\documentclass{iau}

\usepackage{float}
\usepackage{amsmath}
\usepackage{graphicx}
\usepackage{multirow}
\usepackage{authblk}
\newcommand{\bit}[1]{\ensuremath{\textit{\bfseries{#1}}}}

\begin{document}

\lefttitle{R. Narasimha, M. Safonova, C. Sivaram}
\righttitle{Making Habitable Worlds: Planets Versus Megastructures}

\jnlPage{1}{7}
\jnlDoiYr{2021}
\doival{10.1017/xxxxx}

\aopheadtitle{Kavli-IAU Symposium (IAUS 387)}

\title{Making Habitable Worlds: Planets Versus Megastructures}

\author[1,2]{Raghav Narasimha}
\author[2]{Margarita Safonova}
\author[2]{Chandra Sivaram}

\affil[1]{Christ University, Bangalore, India}
\affil[2]{Indian Institute of Astrophysics, Bangalore, India}

\begin{abstract}
With billions of planets in the galaxy, advanced civilizations could relocate planets within or into their planetary systems rather than destroy entire planetary systems to construct megastructures. Such shifts could create Strange Exoplanetary Architectures (SEA), with unusual planetary arrangements potentially indicating deliberate actions by extraterrestrial intelligence (ETI). Searching for biosignatures and technosignatures in these systems could be a promising method for detecting ETI activities.
\end{abstract}


\maketitle

\section{Introduction}

Star KIC~8462852\footnote{Formally Boyajian's star, now known as Tabby's star after its discoverer Tabetha Boyajian.} caught the world's attention when the observers \citep{Boyajian_2016} struggled to explain the irregular dips in the brightness. These unusual dips could not be explained by stellar variability or planetary transits. Theories involving swarms of comets or asteroids proved inadequate, thereby propelling an alternative hypothesis into the forefront -- that KIC~8462852 could be enveloped by a Dyson Sphere (DS) constructed by an advanced ETI to harness the star's energy output, and this is what causing these variations \citep{Wright_2016}. This has put the world into an observational frenzy centred on this star. Finally in 2018, the light curve made in three colours showed chromaticity -- the blue light was blocked much more than red light during the star's dimming, a characteristic of the dust (from comets or asteroids) rather than of something made from solid material \citep{Boyajian_2018,Deeg_2018,Schaefer_2018}, but it was to late to thwart ET hopes -- the world was woken up to the idea of megastructures.

The possibility of building giant structures by advanced civilization to better utilize the energy output from their star was suggested very early in the 20th century, possibly first in a 1929 non-fiction book by J.~D.~Bernal, which describes the space habitats intended for permanent residence in case Earth becomes unlivable, able to host several hundred thousand people \citep{bernal_1929}. The knowledge of the Solar System was well developed at that time (Pluto was discovered in 1930), and Bernal was suggesting using asteroids, Saturn's rings, etc., to construct his habitation globes: ``{\it \dots the great bulk of the structure would be made out of the substance of one or more smaller asteroids, rings of Saturn or other planetary detritus.}". Olaf Stapledon further developed the concept of space habitats in his 1937 book {\it The Star Maker}, where he described that stars without natural planets can be made habitable by constructing artificial habitats around them \citep{Stapledon_1937}. These artificial worldlets -- hollow globes of different sizes -- were to be arranged in concentric rings at some particular distance from the star. All this was possibly the inspiration for Dyson to introduce his megastructure. Dyson acknowledged that Stapledon’s conceptualization of an artificial biosphere served as an inspiration for his own work \citep{Dyson_1979}.

\section{Dyson Sphere}\label{sec-ds}

The Search for Extraterrestrial Intelligence (SETI) enterprise originated from Morrison and Cocconi's suggestion to search for radio signals emitted by ETI \citep{Cocconi_1959}. In 1960, Freeman Dyson proposed an alternative way of detecting ET technology -- by looking for sources in the infrared (IR) region that are radiating at temperatures suitable for sustaining life, yet with a power output equivalent to that of a star \citep{Dyson_1960}. Dyson's conceptualization of his artificial biosphere (megastructure) was based on three key assumptions:
\begin{itemize}
\it
   \item[-] Malthusian theory of population growth. 
   \item[-] Our own Solar System as a model for other planetary systems. 
   \item[-] Jupiter as the resource for building the megastructure.
\end{itemize}
\noindent
Dyson hypothesized that if the total mass of Jupiter were evenly distributed within a spherical shell around the Sun at an orbital radius of approximately 1 astronomical units (AU), the resulting shell -- with a thickness of 2 to 3 meters and a mass\footnote{Expressed here in mass per unit area, because of unknown construction material.} of 200 g/cm$^2$. A shell of such thickness could be made comfortably habitable and can contain all machinery required for utilizing all solar radiation falling on it creating an artificial biosphere around the star. The temperature of such megastructure might be around 200 to 300 K, which would be radiating at a wavelength around 10 micron \citep{Dyson_1960}. Dyson further expounded his idea by saying that \citep{Dyson_1966}\footnote{The emphasis is added by authors.}:
\begin{itemize}
\it
 \item[-] It is possible to build \bit{large} rigid structures in space.
 \item[-]  It is possible to build \bit{light} rigid structures in space.
 \item[-] It is possible to take planets apart.
\end{itemize}

\section{Drawbacks of Dyson's Idea}\label{sec-db}

\subsection{Wrong assumptions}

{\it A. Malthusian theory of population growth:}

The Malthusian theory, which posits that population growth will inevitably outstrip the availability of resources, has faced considerable criticism due to its inability to account for technological advancements, shifts in societal norms, and other factors that have influenced population dynamics. It has been shown to be flawed in various respects \citep{malthus_2019}. Dyson himself acknowledged that his assumption of population growth surpassing the Malthusian limit was "anthropomorphic" \citep{Maddox_1960}.

{\it B. Our own Solar System as a model for other planetary systems:} 

Dyson wrote this article over three decades prior to the discovery of exoplanets; he must have thought that our Solar System is a typical planetary system. In fact, most known planetary systems do not have enough planets (rocky construction material) to build megastructures; the majority of known planets are in single-component systems; only one more system is known, apart from the solar system, to host eight planets (KOI-351).

{\it C. Jupiter as the resource for building the megastructure:}

Only about 13\% of Jupiter's mass is practically usable for construction, as Jupiter is primarily composed of hydrogen and helium \citep{Militzer_2008}. Even if we were to break up all the inner planets, we would only obtain $11.79\times10^{24}$ kg of material, which is sufficient for a 1-AU shell with a mass of 42 kg/m$^2$, equating to a mere 1.5 cm in thickness. If we used all the rocky material available in the Solar System (approximately $1.82\times10^{26}$ kg, including the gas giants' cores), it would allow for a 1-AU shell with an average thickness of 8 cm, resulting in a mass of 650 kg/m$^2$ \citep{harrop_2010}. Dyson himself clarified that “a solid shell or ring surrounding a star is mechanically impossible." What Dyson actually envisioned was a “collection or swarm of objects traveling on independent orbits around the star" \citep{Maddox_1960}.

\subsection{Destruction of Heliosphere}

Solar System's heliosphere serves as a protecting shield for life on our planet, shielding more than 75\% of cosmic rays from the Milky Way galaxy \citep{Stone_2016}. Constructing a DS around the Sun will trap the solar wind inside, resulting in the obliteration of the heliosphere. This, in turn, would allow the interstellar medium and cosmic rays to affect the DS. Whether the biosphere is inside or outside the DS, it will be affected. Even shrinking of the heliosphere can affect life on the Earth. Notably, even a reduction in the size of the heliosphere can exert profound effects on life on the Earth. A historical instance exemplifies this: approximately 2 to 3 million years ago, the heliosphere experienced a contraction, shrinking to a mere 0.22 AU  in radius. Coinciding with this event was one of the episodes of mass extinction that occurred during the same period \citep{Opher_2022_1, Opher_2022_2}. This correlation underscores the intricate relationship between the state of the heliosphere and the sustenance of life on Earth, emphasizing the significance of its role as a protective boundary against cosmic hazards.

\section{An Alternative: Moving Planets into the Habitable Zone}\label{move}

\noindent
The motivation behind constructing megastructures stemmed from the aim of addressing the energy crisis faced by advanced overpopulated civilisation -- with the eventual need for extra living space and resources.  It is worth considering that we can potentially tackle the energy crisis by delving into aspects that Dyson might not have accounted for initially. Dyson's premise hinged on the idea of utilizing Jupiter as a resource, albeit without acknowledging a critical factor: only a fraction of Jupiter's mass, approximately 13\%, is practically employable for construction purposes due to the predominance of hydrogen and helium in its composition. 

If the energy requirement of some advanced civilisation increases, they even might have advanced commercial nuclear fusion reactors at their homes to fulfill their energy needs (like how we have solar panels right now) instead of building DS. Even if the energy sources run out on the Earth because of high energy demands, the gas giants are there for the rescue! If we convert Jupiter's hydrogen content to energy by thermonuclear reactions, the energy released is $\sim 10^{42}$ J. Even if we consume energy at the rate of solar luminosity, we can manage for more than 100 million years \citep{Carl_1966}. However, as the global population increases, the urgency for more habitable area and resources intensifies, necessitating an expansion beyond our planet. Basically, we may need more habitable planets!

It is theoretically possible to manipulate planetary trajectories using lasers. By deploying multiple lasers from different directions, it is feasible to manage the transfer of momentum, thereby enabling precise adjustments to a planet's trajectory while minimizing unintended effects on other planets. Any displacement of a planet should be done very gradually to minimize the disruptions to the orbital dynamics of the system. Sudden changes could have unpredictable effects, potentially destabilizing other orbits. The use of laser technology to reposition planets in the habitable zone (HZ) would facilitate the terraforming process. The notion of moving planets existed more than a century ago. In 1895, Konstantin Tsiolkovsky suggested that the intelligent beings of the future may control the motion of the planets, saying ``{\it They drove the planet as we drive horses}" \citep{tsiolkovsky_1895}.

Shifting planets like Mars, Pluto, or free-floating planets into the Solar System's HZ  could significantly alter their environments, potentially making them more suitable for life. For Mars, moving it closer to the Sun might warm the planet, melting polar ice caps and releasing water, which could aid in terraforming efforts. However, the increased solar radiation could also cause atmospheric changes, potentially leading to a greenhouse effect or atmospheric loss. Pluto, if relocated to the HZ, would experience a substantial rise in temperature, requiring atmospheric and surface transformations to support life. Free-floating planets brought into the HZ could also undergo similar changes, possibly developing habitable conditions if they possess a sufficient atmosphere and water. However, the challenges and risks associated with such endeavors would require careful planning and advanced technology. Even Trans-Neptunian objects (TNO) and asteroids, when moved into the HZ, could become sources of water or other resources and possibly develop more dynamic environments.

ETI might intentionally relocate planets within their system for purposes like industrial exploitation, energy generation, or waste processing. These uninhabitable planets, termed "service worlds" \citep{Wright_2022}, could serve various functions according to their characteristics. For example, gas giants might be moved closer to harness their hydrogen for nuclear fusion, while icy planets like Pluto could support aquatic life or aquaculture. Rocky planets could be used for large-scale agriculture or even as astronomical observatories.

\section{Energetics involved in moving planets into the HZ}\label{energy}

\subsection{Hohmann Transfer}

\begin{figure}[h]
\centering
\includegraphics[width=0.475\textwidth]{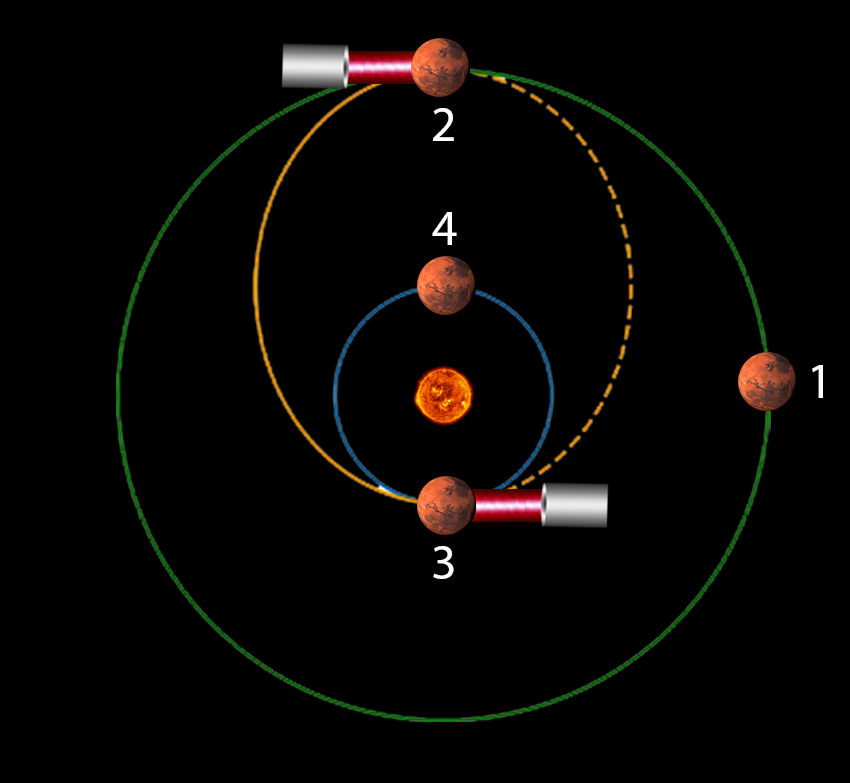}
\caption{Hohmann transfer of Mars using lasers (the image is not to scale). The representation of a single large laser is merely illustrative; in practice, the configuration might involve multiple lasers operating from various orientations.}\label{fig-hohmann}
\end{figure}

The Hohmann Transfer (HT) is an efficient method to move a smaller object between orbits around a larger body, like the Sun (See Fig.~\ref{fig-hohmann}), using minimal energy by applying two velocity changes \citep{sivaram_2008}. Shifting Mars $0.3$ AU into the HZ requires slowing it by $1.3$ km/s at $1.5$ AU and $1.4$ km/s at $1.2$ AU, using $\sim 10^{32}$ J and taking $\sim 10$ months. Moving Pluto $37.8$ AU into the HZ requires slowing it by $3.5$ km/s at $39.5$ AU and 10.7 km/s at $1.2$ AU, using $\sim 10^{29}$ J and taking $\sim 47$ years. While HT is efficient, slower transfers may be preferable for maintaining planetary stability.

\subsection{Laser Power for Continuous Thrust}

We assume the laser used to shift a planet operates at constant power, positioned close enough to the planet that beam divergence is negligible. The equation of motion for a constant-power laser is as follows:

$$ \frac{d^2 r}{dt^2} = \frac{P}{M_p \cdot c}, $$

where $r$ is the distance to be moved, $P$  is laser power, $M_p$ is the mass of the planet and $c$ is the speed of light. The required laser power is:

$$ P = \frac{M_p \cdot r \cdot c}{t^2}. $$

For example, shifting a Mars-sized planet $0.3$ AU into the HZ over $10^{11}$ seconds (few thousand years) requires $\sim 10^{21}$ W and $\sim 10^{32}$ J. Similar energy levels apply for moving Pluto- or Earth-sized planets, while shifting asteroids would require significantly less energy.

It is suggested that power-beaming technology can be used for launching spacecraft, raising satellite orbits, facilitating space-to-space transfers, and even for missions to the outer solar system and beyond \citep{benford_2016}. Power beaming could potentially be extended to include the displacement of asteroids and planets, which would require significantly higher orders of power.

\section{Detection methods}\label{sec-dm}

Searches for ETI in radio, infrared, and optical regions have formed the basis of modern SETI \citep{shuch_2011}. However, despite decades of our efforts, concrete evidence of ETI remains elusive. The null result does not mean that there are no ETI. ``{\it Our current search completeness is extremely low, similar to having searched for something like a large hot tub or a small pool’s worth of water out of all of Earth’s oceans}" \citep{Wright_2018}. 

Continuous improvement of search parameters and models is essential, along with exploration of novel detection ideas. If advanced civilizations use high-power lasers to shift planets, this would create detectable technosignatures. These sustained laser signals, or High-Power Laser Technosignatures (HPLT), could be sought, especially in systems with unusual planetary arrangements, which might indicate ETI manipulation.

\begin{figure}[h]
\centering
\includegraphics[width=0.475\textwidth]{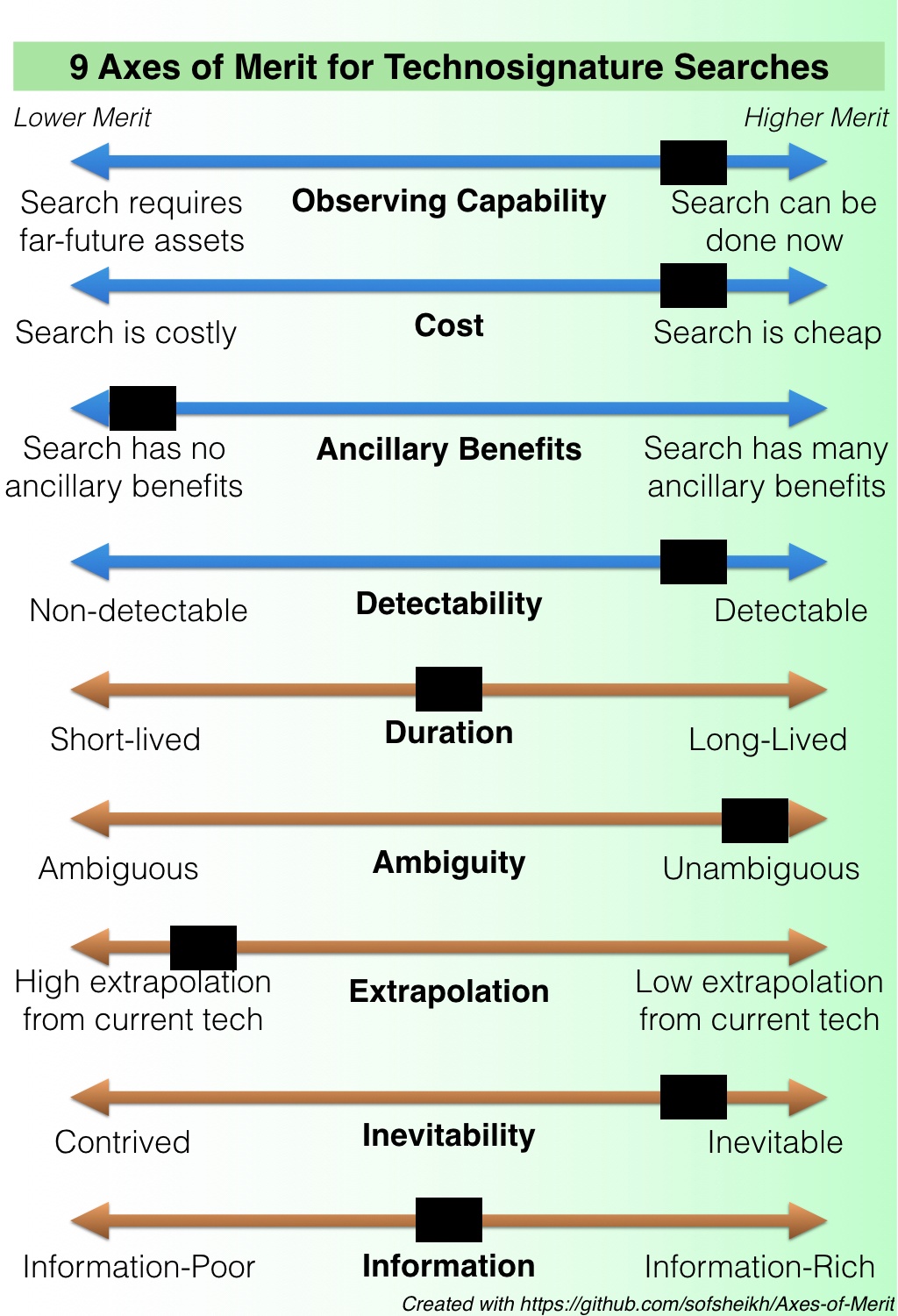}
\caption{Axes of Merit for high-power laser technosignature.}
\label{fig-aom}
\end{figure}

Sofia Sheikh created a framework to rank technosignatures based on nine factors \citep{Sheikh_2019}. This framework allows people to critically evaluate technosignatures on the basis of detection probability. We considered the HPLT and ranked it by the following nine parameters. Here is the ranking and the rationale behind the ranking:
\begin{itemize}

    \item\bit{Observing Capability} We can detect shifting of planets using powerful lasers with our current observing capabilities. No need to build any new instruments for observation.
    \vskip -0.2in
    \item\bit{Cost} The cost of searching for laser signatures is relatively low as no new instrumentation is required.
    \item\bit{Ancillary Benefits} HPLT detection has no present ancillary benefits.
    \item\bit{Detectability} HPLT are detectable even over a kiloparsec.
    \item\bit{Duration} HPLT are not transient in nature as they operate continuously over years. The time period of the high power laser signature depends upon the distance to which the planet need to be moved. We estimate that if the acceleration is low (like $10^{-10}$ m/s$^2$), then it may take around 1000 years to move a Earth-sized planet over~0.5 AU, and even more time is required to shift a FFP from interstellar space to the resident star system.  
    \item\bit{Ambiguity} The high-power narrow-band laser beam has no natural source. Therefore, these technosignatures are unambiguous.
    \item\bit{Extrapolation} High-power lasers (at ZW scale) operating continuously are yet beyond our current technological capabilities, but we already have such powerful lasers operating at short pulses. Thus, building such high-power lasers operating in continuous mode is plausible in future.
    \item\bit{Inevitability} The need for additional living space and resources is inevitable. Therefore, additional habitable planets could eventually be absolutely necessary for advancing and space-faring civilizations. 
    \item\bit{Information} ETI can also transmit information through their lasers and the information is proportional to the bandwidth of the laser. 
\end{itemize}
These merits are visually represented in Fig.~\ref{fig-aom}, with Low to High merits displayed from left to right.

If an advanced species had already relocated planets into their HZ, we might expect to observe an unusually high concentration of planets in close proximity and SEAs. For example, TRAPPIST-1 has four potentially habitable planets, while Kepler-20 shows strange orbital patterns that suggest possible artificial manipulation. A recent discovery of the HD~110067 planetary system \citep{6planets} has reinforced our proposal to look for SEA systems. The star is an old K0-type star with mass and radius of $\sim 80\%$ of the Sun hosting 6 sub-Neptune planets orbiting in a stable perfect resonance, which means that all 6 of them align every few orbits, or 492 days. Since all planets are in the star's HZ and three of the planets have low-densities with atmospheres, they are already considered to be potentially habitable, prompting the recent search for radio technosignatures in this system \citep{Choza2024}. It is plausible that these planetary systems were artificially constructed through the intentional relocation of planets, allowing inhabitants to facilitate travel or transport resources more efficiently between them. Therefore, we suggest that it is advantageous to look for biosignatures and other technosignatures in such engineered systems.

\section{Conclusion}

Building on the ideas of pioneers like Tsiolkovsky, Dyson, and Sagan, there is a growing consensus that humanity may eventually need to leave Earth and expand into space. Given the rapid increase in population and resource consumption over the past century, space colonization may become inevitable. This could be achieved either by constructing megastructures, such as Dyson Spheres, or by shifting planets into the habitable zone (HZ) of their stars. We critically examined the drawbacks of constructing Dyson Spheres and propose relocating planets using high-power lasers as a viable alternative. This approach could also be applied to moving asteroids or Trans-Neptunian Objects (TNOs) with even less energy. High-power beamed lasers used in these operations could be detectable over interstellar distances, making them ideal technosignatures. We recommend searching for biosignatures and technosignatures in planetary systems with significant orbital adjustments, as these systems might have been intentionally engineered.

\end{document}